\documentclass[twocolumn,showpacs,preprintnumbers,amsmath,amssymb,aps]{revtex4}
\usepackage{graphicx}
\begin{document}

\title{Devil's Staircase and Disordering Transitions in 
Sliding Vortices and Wigner Crystals on Random Substrates with Transverse Driving}
\author{C. Reichhardt and  C.J. Olson Reichhardt 
}
\affiliation{  
Theoretical Division,
Los Alamos National Laboratory, Los Alamos, New Mexico 87545}  
\date{\today}

\begin{abstract}
Using numerical simulations we show that, 
in the presence of random quenched disorder, 
sliding superconducting vortices and
Wigner crystals pass through a variety of dynamical phases 
when an additional transverse driving force is applied. 
If the disorder is weak, the driven particles form a moving lattice and
the transverse response shows a devil's staircase structure 
as the net driving force vector 
locks with the symmetry directions of the moving lattice, 
in agreement with the predictions of
Le Doussal and Giamarchi. 
For strong disorder, and particularly for smoothly varying potential
landscapes, the transverse response consists of
a sequence of disordering transitions 
with an intervening formation of stable channel structures.   
\end{abstract}
\maketitle

\section{Introduction}

A number of systems, including moving vortex lattices
in type-II superconductors with disordered pinning 
\cite{Koshelev,Giamarchi,Balents}, 
sliding charge-density waves \cite{Markovic},
and driven Wigner crystals \cite{Noise}, can be modeled as moving lattices 
interacting with random quenched disorder. 
If the external driving force is
weak the particles are pinned, while at high external 
drives the particles can slide. There is extensive evidence
that in the sliding phase  
the particles organize into a moving anisotropic crystal 
or moving smectic phase and 
that the motion is confined 
to well defined channels 
\cite{Koshelev,Giamarchi,Balents,Moon,Olson,Fangohr,Kes}. 
A remarkable property of these channels is that if an additional  
external drive is applied in the {\it transverse} direction the moving 
system exhibits a finite transverse depinning threshold \cite{Giamarchi}. 
This transverse depinning threshold has been
confirmed in a number of simulations \cite{Noise,Moon,Olson,Fangohr,Soret} and 
experiments \cite{Kes}. 

A transverse depinning barrier is also present for particles
moving over a periodic substrate.
In this type of system,   
the transverse velocity force curves exhibit
a devil's staircase structure as the particle motion locks to the 
symmetry directions of the periodic substrate \cite{Reichhardt,Granto,Korda}. 
This effect occurs even for a single particle moving over a 
periodic substrate 
\cite{Grier,Locking,Lacatsa}. 
In contrast, a single particle moving over a random substrate 
will not show a transverse depinning threshold or a 
devil's staircase response. 

In Ref.~\cite{Giamarchi},
a finite transverse depinning force was predicted
for a vortex lattice, and 
it was also conjectured
that a moving lattice would align with the direction of the 
initial longitudinal drive in order
to minimize power dissipation. If the moving lattice remains aligned in 
this initial orientation
as a transverse drive is applied, the transverse velocity 
force curves could exhibit a devil's staircase response as the  
applied force vector locks with the
triangular symmetry of the moving lattice. 
In contrast to a system with a periodic substrate,
here the symmetry breaking which leads to the 
preferred directions of motion comes from the
moving lattice itself. 
Numerically observing 
the devil's staircase in a system with random disorder can be difficult 
since large systems and long time averages of the 
transverse velocity are needed to reduce the fluctuations. 
An added difficultly with observing this type of devil's staircase is that  
fine increments of the applied transverse force
are required to resolve the higher order steps.    

Another possibility for the dynamics at the transverse depinning 
transition which has not 
been considered is that
the moving lattice structure could become strongly 
disordered at the depinning transition, leading to plastic distortions. 
If such a {\it plastic transverse depinning} 
occurs, it is natural to ask if the
particles would reorder into a tilted channel structure, causing a new 
transverse barrier to develop.

\section{Simulation}

\subsection{Vortex System} 

To address these issues we consider two types of 
two-dimensional (2D) systems with periodic boundary conditions. 
The first is a vortex system containing $N_v$ vortices and
$N_p$ short range random pinning sites. 
The vortices are modeled as point particles and
their dynamics evolve via an overdamped equation of motion,
given for vortex $i$ as
\begin{equation} 
\eta \frac{d{\bf R}_{i}}{dt} = {\bf F}_{i}^{r} + {\bf F}^{p}_{i} + {\bf F}_{d},
\end{equation} 
where $\eta$ is the damping constant.
The vortex-vortex interaction force \cite{DeGennes,Tinkham} is  
${\bf F}_{i}^{r} 
= \sum_{i \neq j}^{N_v} f_{0}K_{1}(r_{ij}/\lambda){\bf {\hat r}}_{ij}$,
where $K_{1}(r)$ is the modified Bessel function, 
$r_{ij}=|{\bf r}_i-{\bf r}_j|$ is the distance between vortices $i$ and $j$
located at ${\bf r}_i$ and ${\bf r}_j$,
${\bf {\hat r}}_{ij}=({\bf r}_i-{\bf r}_j)/r_{ij}$,
the unit of force is $f_{0} = \Phi_{0}^2/2\pi\mu_{0}\lambda^3$,
$\Phi_{0}$ is the flux quantum, and 
the London penetration depth $\lambda$ is the unit of length.
This interaction is appropriate for stiff 3D vortices 
in relatively thick films where the film thickness $d$ is much
larger than $\lambda$ but is smaller than the Larkin-Ovchinnikov
correlation length \cite{Larkin,Brass}.
Time is measured in units of $\tau=\eta/f_0$ and a molecular dynamics (MD)
time step is equal to $0.02\tau$.
We find identical results if smaller time steps are used.
For a 0.1$\mu$m thick crystal of NbSe$_{2}$, $f_{0} = 6.78\times 10^{-5}$N/m
and $\eta=2.36\times 10^{-11}$ Ns/m \cite{Karapetrov}, so that
$\tau=0.35\mu$s.
The pinning force ${\bf F}^{p}_{i}$ arises
from pinning sites which are modeled as 
attractive parabolic potentials of radius $r_{p}$
and strength $f_{p}$:
${\bf F}_i^{p}=-\sum_{k=1}^{N_p}f_p(r_{ik}/r_p)
\Theta(r_p-r_{ik}){\bf r}_{ik}$,
where $r_{ik}=|{\bf r}_i-{\bf r}_k|$ is the distance between vortex $i$ and
pin $k$ located at ${\bf r}_k$, 
${\bf {\hat r}}_{ik}=({\bf r}_i-{\bf r}_k)/r_{ik}$,
and $\Theta$ is the Heaviside step function.
The pinning interaction is intended to represent 
columnar pinning \cite{Bose}.   
The net external driving force ${\bf F}_{d}$ is comprised of 
the longitudinal force applied in the $x$-direction, 
${\bf F}^{L}_{d}  = F^{L}_{d}{\bf {\hat x}}$, 
and the transverse force applied in
the $y$-direction, ${\bf F}^{Tr}_{d} = F^{Tr}_{d}{\bf {\hat y}}$. 
The vortex (pin) density is given by $\rho_v$ ($\rho_p$). 
The system is prepared by simulated annealing, after which a longitudinal
force is gradually applied
in increments of
$1\times 10^{-4}f_0$ up to a  specific value. 
The final longitudinal driving force is held
fixed and then the transverse drive is applied in small increments. 
We measure the transverse velocity
$V^{Tr}=\langle N_v^{-1}\sum_{i=1}^{N_v} {\bf v}_i \cdot {\bf {\hat y}}
\rangle$,
where ${\bf v}_i$ is the velocity of vortex $i$.
We also measure the fraction of sixfold coordinated vortices,
$P_6=N_v^{-1}\sum_{i=1}^{N_v}\delta(6-z_i)$, where $z_i$ is the
coordination number of vortex $i$ as determined from a Voronoi
construction.

\subsection{Sliding Wigner Crystals} 

The second system we consider is sliding Wigner crystals
composed of $N_e$ electrons.  Here we use
the same Langevin type simulations which were
employed previously to identify the sliding
states in this system \cite{Noise}. 
The 
electron-electron interaction force 
${\bf F}_{i}^{r}=-\nabla U_i^{r}$, where
\begin{equation} 
U_{i}^{r} = \sum_{i \neq j}^{N_e}
\frac{e^{2}}{{|{\bf r}_{i} - {\bf r}_{j}|}}
\end{equation} 
and ${\bf r}_{i(j)}$ is the location of election $i(j)$. The 
electron crystal has lattice constant $a_{0}$. 
The disorder comes 
from $N_p$ positively charged impurities placed at a distance 
$d=0.9a_0$ out of plane with the form   
${\bf F}_{i}^{p}=-\nabla U_i^{p}$, where
\begin{equation} 
U_{i}^{p}=-\sum_{j}^{N_p} \frac{e^{2}}
{\sqrt{|{\bf r}_{i} - {\bf r}^{(p)}_{j}|^{2} + d^{2}}}. 
\end{equation}
Here  ${\bf r}_{j}^{(p)}$ is the in-plane projection of the location
of pin $j$. 
The number of electrons $N_e$ 
equals the number of impurities $N_p$.
Both the electron-electron and electron impurity interactions are long 
range, and we evaluate them with a fast-converging sum \cite{Niels}.
To measure the transverse response, 
we follow a similar procedure as in the vortex system. 

\section{Transverse Depinning For Vortex Systems}   
\subsection{Definition of Pinning Regimes}

We first consider a vortex system
for a sample of size $L = 24\lambda$ 
with $\rho_v=\rho_p=0.73/\lambda^2$,
$r_{p} =  0.2\lambda$, and   
fixed longitudinal drive $F^{L}_{d} = 0.012f_0$. 
For this value of $F^{L}_{d}$ and pinning density we find three regimes  
of the transverse response for varied pinning strength. 
For $f_{p} < 0.085f_0$ the system is in what we term the weak pinning limit, 
and
the equilibrium state at $F^{L}_{d} = 0.0$ 
is free of dislocations.
At $F^{L}_{d} = 0.012f_{0}$, 
there is a finite transverse depinning threshold and the
transverse depinning occurs without the generation of
additional defects. 
In the intermediate regime $0.085 \leq f_{p}/f_{0} < 0.13$,
the equilibrium $F_d^L=0$ state contains dislocations, and
for $F^L_d=0.012f_0$ 
the transverse depinning 
is plastic and is accompanied by the generation of defects. 
Under a longitudinal drive, for 
$0.085 \leq f_{p}/f_{0} < 0.106$ 
the vortices
dynamically reorder into a defect-free lattice, while for
$0.106 \leq f_{p}/f_{0} < 0.13$ 
the vortices undergo a partial dynamical reordering 
into a moving
smectic state containing a finite number of dislocations that are all
aligned with the driving direction, as seen in previous simulations
\cite{Moon,Olson}.
The number of defects in the
moving smectic state gradually increases with increasing $f_{p}$ until each 
row of vortices is completely decoupled from the neighboring rows.
A weak smectic state contains only a few dislocations 
so that several adjacent rows are coupled but rows separated by longer
length scales are decoupled.
In the strong pinning regime $f_{p} \ge 0.13f_0$, the vortices
flow plastically through the system and the
moving channel structure is destroyed 
so that the dislocations are no
longer aligned with the drive. 
In this case there is no transverse depinning barrier.   
We have tested system sizes up to $L=92\lambda$ and we find only
dislocation free lattices in the equilibrium state for $f_p<0.085f_0$.
It is possible that for much larger systems than we can simulate,
dislocations will eventually appear in the weak pinning systems.

\begin{figure}
\includegraphics[width=3.5in]{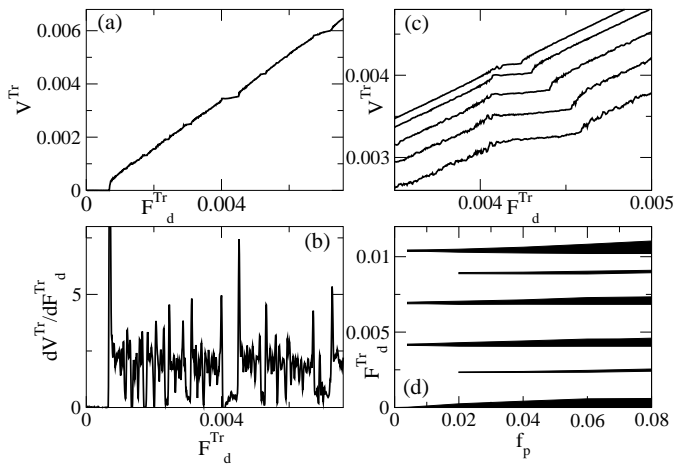}
\caption{
(a) The transverse velocity 
$V^{Tr}$ vs the transverse applied force $F^{Tr}_{d}$  
for a vortex lattice moving over random disorder
of strength $f_p=0.06f_0$
with fixed longitudinal driving force $F^{L}_{d}=0.012f_0$.
(b) $dV^{Tr}/dF^{Tr}_{d}$ vs $F^{Tr}_{d}$ for the same system.
(c) $V^{Tr}$ vs $F^{Tr}_{d}$ showing the evolution of  the
$(1,2)$ step for $f_{p} = 0.02$, 0.04, 0.06, and $0.08f_0$, 
from top to bottom.
(d) Step widths $w_s$ in $F^{Tr}_d$ vs $f_{p}$ 
for the $(0,0)$, $(1,3)$, $(1,2)$, $(3,4)$, $(3,3)$ 
and $(4,4)$ steps, from bottom to top.}   
\label{fig:image}
\end{figure}

\subsection{Elastic Transverse Depinning} 

We consider a sample with $f_p=0.06f_0$ in the weak pinning regime where the 
equilibrium lattice contains no dislocations.
We fix the longitudinal drive to $F^{L}_{d} = 0.012f_{0}$  
and increase $F^{Tr}_{d}$ from zero 
in increments of $4\times 10^{-5}f_0$ every $1.5\times 10^{4}$ 
MD time steps.    
In order to resolve the steps in the transverse response, 
we average the transverse velocity $V^{Tr}$ 
over $9\times 10^{4}$ MD steps. 
The appearance of the curve does not change if
$V^{Tr}$ is averaged over $1.5 \times 10^{4}$ MD steps instead.
A typical simulation runs for a total of more than $5\times 10^8$ MD steps.  
When the net driving force vector aligns with a symmetry direction of the 
vortex lattice, we find that the transverse response 
locks to that direction over a range of $F^{Tr}_{d}$.
For a moving triangular lattice, this locking
occurs whenever $F^{Tr}_{d}/F^{L}_{d} = \sqrt{3}m/(2n + 1)$ where 
$n$ and $m$ are integers.  In such a devil's staircase, 
the most pronounced steps occur
when both $n$ and $m$ have low values.  

The plot of $V^{Tr}$ vs $F^{Tr}_{d}$ 
in 
Fig.~\ref{fig:image}(a) 
shows that a transverse critical depinning threshold $F^{Tr}_c$ occurs at 
$F^{Tr}_{d} = 0.00065f_0$. 
It is important to note that the vortex lattice 
remains aligned in the $x$-direction during the entire simulation. 
For increasing $F^{Tr}_{d}$, 
a number of steps appear in $V^{Tr}$ corresponding to the locking of the
net applied force vector with a
symmetry direction of the lattice. A prominent step occurs 
near $F^{Tr}_{d} = 0.0043f_0$ which 
corresponds to the case of $(m,n) =  (1,2)$.
The step at $F^{Tr}_{d} = 0.003f_0$ corresponds to 
the (1,3) state, and the (2,3) state appears at $F^{Tr}_{d} = 0.007f_0$. 
There are also a number of smaller steps at higher values of $n$ and $m$, 
confirming the existence of
a devil's staircase response. 
In Fig.~\ref{fig:image}(b), $dV^{Tr}/dF^{Tr}_{d}$ vs $F^{Tr}_{d}$
more succinctly  
highlights the steps. The peaks in Fig.~1(b) correspond to 
points at which the curvature
in $V^{Tr}$ vs $F^{Tr}_{d}$ increases just before and after a plateau region.
The most pronounced steps occur for $(m,n)$ where the value of both $m$
and $n$ are small, which is typical for a devil's staircase structure. 
This result confirms the prediction 
of Giamarchi and Le Doussal \cite{Giamarchi} 
that if the moving lattice stays aligned with the original
applied longitudinal drive, a devil's staircase response 
occurs due to the matching
of the symmetry directions of the lattice with the applied drive.   

The width of a given step $w_s$ depends on the pinning strength and density. 
In Fig.~\ref{fig:image}(c) we illustrate
the evolution of the $(1,2)$ step for $f_{p}/f_0 = 0.02$, 0.04, 0.06, 
and $0.08$,
showing the growth of $w_s$ with $f_{p}$. 
We also find that the step widths increase when the pinning density 
$\rho_{p}$
is increased. 
In Fig.~\ref{fig:image}(d) we plot the evolution of $w_s$ 
as a function of pinning strength for $(m,n) = (0,0)$,
$(1,3)$, $(1,2)$ 
$(3,4)$, $(3,3)$ and  $(4,4)$.  
We have verified that these results are robust up 
to systems of size $L = 60\lambda$ 
provided that the lattice remains free of dislocations.
In two-dimensional pinned systems
it has been argued that 
dislocations eventually appear at large distances
even for weak disorder.  These
dislocations would limit the resolution of the devil's staircase,  
which should remain
observable as long as the distance between dislocations
is comparable to or greater than the system size.

\begin{figure}
\includegraphics[width=3.5in]{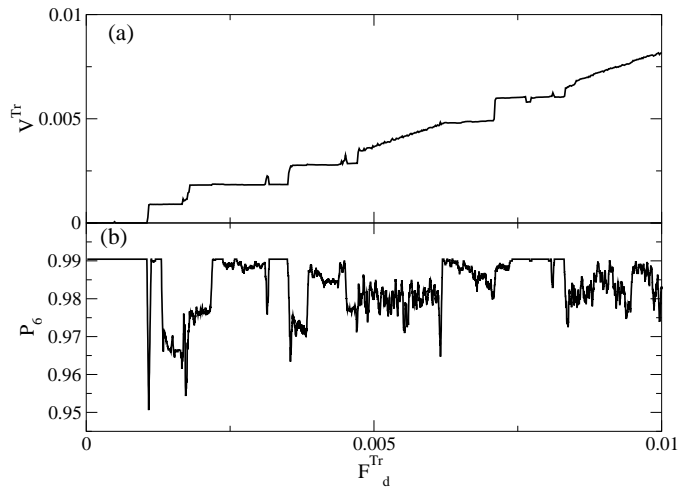}
\caption{
(a) $V^{Tr}$ vs $F^{Tr}_{d}$ 
for a moving vortex lattice with the same parameters as
in Fig.~1(a), but with $f_{p}= 0.106f_0$. 
(b) The fraction of six-fold coordinated
vortices $P_{6}$ vs $F^{Tr}_{d}$ 
for the same sample showing the correlations of 
the drops in $P_{6}$ with the transitions in $V^{Tr}$.   
}
\label{fig:iv}
\end{figure}

\subsection{Plastic Transverse Depinning} 

\begin{figure}
\includegraphics[width=3.3in]{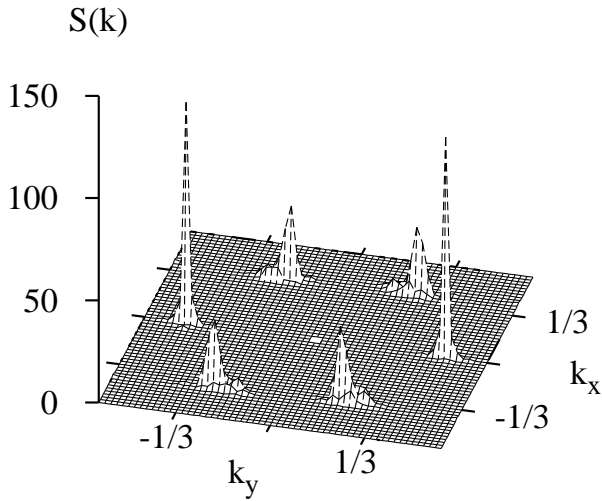}
\caption{
Structure factor $S({\bf k})$ for a system with $f_p=0.106f_0$
in the moving smectic state at $F_d^{Tr}=4 \times 10^{-4}f_0$
and $F_d^L=0.012f_0$.  The central peak is not shown.
}
\label{fig:sk}
\end{figure}

In the 
upper range of the
intermediate pinning strength regime $0.106 \le f_p/f_0 < 0.13$,
the longitudinally driven vortices form a moving smectic state and
the transverse depinning transition is {\it plastic.}
Here the transverse velocity response
is characterized by a 
series of sharp disordering-reordering transitions, 
giving rise to consecutive pronounced steps in
$V^{Tr}$, as shown in Fig.~\ref{fig:iv}(a) for a system with $f_p=0.106f_0$. 
This ordering-disordering velocity-force response, 
which is distinct from a devil's staircase structure,
occurs when the pinning is strong enough that the
particles move in channels that are separated by
a number of aligned dislocations,
which is indicative of a smectic ordering. 
We also find two prominent peaks in the
structure factor $S({\bf k})$ consistent with a smectic state, 
as illustrated in Fig.~\ref{fig:sk}.
Here
$S({\bf k}) = N_v^{-1}|\sum^{N_v}_{i=1}\exp(-i{\bf k}\cdot{\bf r}_{i})|^2$.
In addition to the two prominent peaks, Fig.~\ref{fig:sk} shows
the presence of four smaller peaks.  Since 
$f_{p} = 0.106f_{0}$ falls at the edge of the regime where the moving
smectic state forms, there are only a small number of dislocations present
in the moving smectic state and correlations between neighboring channels 
of vortices occur over a width of several channels.
These correlations produce the four smaller peaks
in $S({\bf k})$. For larger system sizes $L$ these 
four small peaks shrink in magnitude while the two prominent 
peaks remain prominent. 
In Ref.~\cite{Moon}, the prominent peaks were observed to decay algebraically
with system size and the state was termed a
moving transverse Bragg glass.
It is beyond the scope of this work to confirm the existence of
a moving transverse Bragg glass. For larger values of
$f_{p}$, the dislocation density in the moving smectic state increases, 
the correlation length between the moving vortex channels decreases 
to the width of a single channel,
and only two peaks appear in $S({\bf k})$. 
For 
$f_{p} > 0.13f_0$ 
the channel structure is destroyed and 
the vortices move plastically.    
 
\begin{figure}
\includegraphics[width=3.5in]{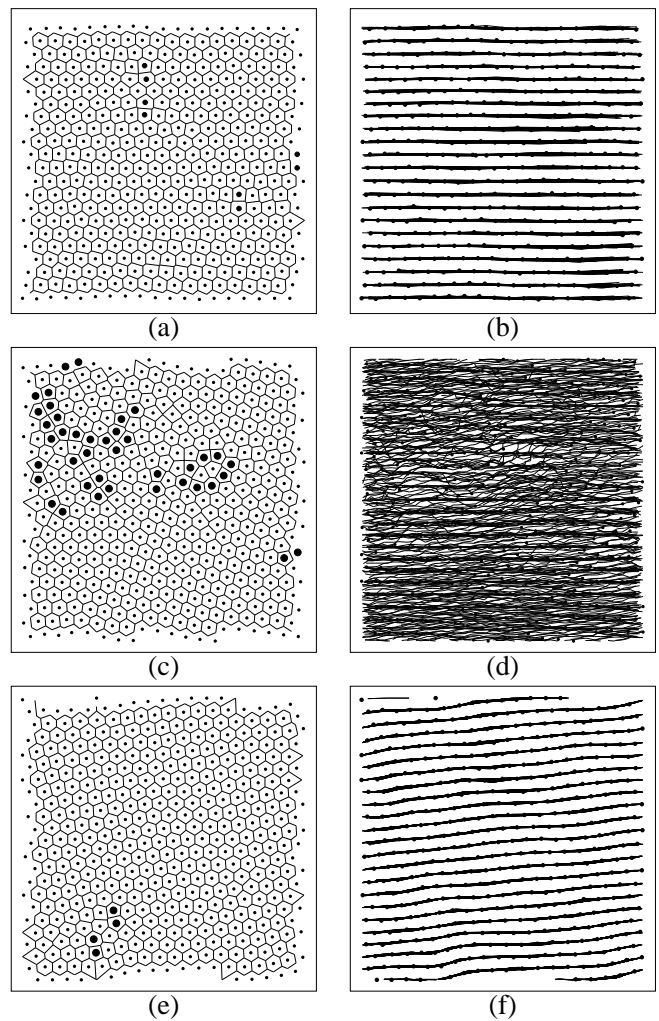}
\caption{
(a,c,e):  Voronoi construction for the vortex configurations 
for the system in Fig.~\ref{fig:iv} taken from a
single snapshot.  Small dots: sixfold-coordinated vortices;
large dots: nonsixfold-coordinated vortices.
(b,d,f): Vortex trajectories (black lines) during a fixed time span.
(a,b): Transversely pinned state at
$F^{Tr}_{d} = 4\times 10^{-4}f_0$ 
where a clear channel
structure appears. (c,d):  The transition state at 
$F^{Tr}_{d} = 0.0011f_{0}$. (e,f): The state at the first 
plateau
in $V^{Tr}$ at $F^{Tr}_{d} = 0.00125f_0$ 
where the system has reorganized into a tilted
channel structure.  
}
\label{fig:delaunay}
\end{figure}

Since the transverse depinning process 
at $f_p=0.106f_0$ is plastic,
a large number of transient dislocations are induced
at the depinning transition which destroy the well ordered
channel structures.
In Fig.~\ref{fig:iv}(b) 
we plot the fraction of sixfold-coordinated 
vortices $P_{6}$ for the same system in Fig.~\ref{fig:iv}(a), indicating that 
the changes in the behavior of $V^{Tr}$ coincide with 
changes in $P_{6}$.  
In the regions where $V^{Tr}$ is increasing linearly 
with drive, $P_{6}$ 
drops
and shows rapid fluctuations.
Here, the 1D channel structure of the moving lattice is destroyed.  
We note that 
for $0.085 \le f_{p}/f_0 < 0.106$ we do not observe any dislocations in the
longitudinal moving lattice; however, since the system is close to 
the value of $f_p$ at which dislocations appear in 
the longitudinally moving lattice, 
at the transverse depinning transition
the extra strain on the lattice is enough to induce some dislocations 
and the system
also depins plastically. 

To further illustrate the change in lattice structure that occurs
on the steps in the intermediate pinning regime,
in Fig.~\ref{fig:delaunay}(a) we show the Voronoi construction 
of the vortex lattice from the system in Fig.~\ref{fig:iv}
for a transversely pinned state
at $F^{Tr}_{d} = 4\times 10^{-4}f_0$.
There are a small number of dislocations present, indicated by large
dark circles, and all of the dislocation pairs are aligned with the
direction of the drive (to the right in the figure).  The vortex
lattice is also aligned with the driving direction.
In Fig.~\ref{fig:delaunay}(b) 
the vortex positions (black dots) and trajectories (black lines)  
indicate that the motion is confined to well defined 1D channels. 
Above the transverse depinning transition, illustrated in
Fig.~\ref{fig:delaunay}(c,d) for $F^{Tr}_{d} =0.0011f_0$,
numerous dislocations are present as indicated by Fig.~\ref{fig:delaunay}(c),
while the trajectories in Fig.~\ref{fig:delaunay}(d) 
show that the channeling effect has been lost.
This disordered type of flow 
appears both at the transitions among the higher steps
as well as in the regions where $V^{Tr}$ is increasing linearly.    

In Fig.~\ref{fig:iv}(a), the velocity jumps to a finite value and 
becomes locked onto a plateau 
for $0.0011f_0 < F^{Tr}_{d} < 0.0017f_0$. 
Here and on the other plateaus in $V^{Tr}$,
the lattice moves in a channel structure which is 
aligned at an angle to the $x$ direction,
such as that shown in Fig.~\ref{fig:delaunay}(e,f) for $F_{d} = 0.00125f_0$. 
As $F^{Tr}_{d}$ increases, the system undergoes repeated 
plastic depinning transitions, each of which appears as a drop
in $P_6$.
In the plastic flow region that follows each of these depinning
transitions,
the vortices move in the
direction of the net force vector.  As the 
transverse drive increases to a new
plateau, the lattice reorders into a new channel structure which
sets up a new transverse barrier to further increases in $V^{Tr}$. 

The channels require some time to reform, so the width of the reordering
transition as a function of transverse driving force 
is dependent on the sweep rate of $F^{Tr}_d$.
For higher sweep rates, the width of the disordered regions between
plateaus grows. In the adiabatic 
regime, the steps would be much sharper, suggesting that the transitions
are first order in nature. We also find hysteresis 
in the steps when we reverse
$F^{Tr}_{d}$. 
For very strong pinning, $f_{p} > 0.13f_0$, and fixed $F^L_d=0.012f_0$,
the channel structures are destroyed and the vortices flow plastically
for all transverse drives; 
however, for higher values of $F^{L}_{d}$, a partially ordered phase appears
and the transverse depinning occurs in the
same manner as in Fig.~\ref{fig:iv}.   
In general, for stronger disorder the longitudinal moving state contains
numerous dislocations
and the transverse depinning occurs in a step like fashion. 

The transition between devil's staircase behavior 
at weak pinning and the order-disorder
transitions associated with intermediate pinning strength
is determined by the appearance of dislocations
in the longitudinally driven lattice.
In general, a devil's staircase 
appears when the disorder is weak enough that there are no dislocations 
present at any drive.
For this reason, no evidence of a devil's staircase appeared in the
simulations of Ref.~\cite{Moon}, which
showed the existence of a transverse depinning barrier.
These simulations were performed in the intermediate pinning limit,
as indicated by the linear scaling of $f_{c}$ with 
the pinning strength in Ref.~\cite{Moon} and the presence of aligned
dislocations in the lattice. 
The ordering-disordering steps that we find in the intermediate pinning limit
were not observed in Ref.~\cite{Moon} since the 
resolution of the transverse force versus velocity curves was too low.
The existence of a 
weak pinning regime may depend on system size. It has been argued
that for any 2D system with random disorder, 
dislocations will always appear at sufficiently long length scales
$R_D$  
even if the disorder is weak.
In our simulations we have considered
numerous different system sizes, 
and 
for 
$f_{p} < 0.085f_0$
we only observe
dislocation free lattices. 
Due to computational constraints we could only simulate systems
of size $L = 92\lambda$ and less. 
It is possible that for much larger systems,
dislocations will eventually appear.
If enough dislocations are present to destroy the orientational order, 
the devil's staircase behavior would be lost when the 
well defined lattice directions of the moving lattice are lost. 
It is then possible that for system sizes
$L<R_{d}$ the devil's staircase behavior will occur, 
while for $L  > R_{d}$, the
strong pinning behavior will be observed. On the other hand, if the
system is in 3D with point pinning but with stiff vortices,
than a 3D moving Bragg glass could form without dislocations, permitting a
devil's staircase behavior to be observed for all lengths.
We also note that the argument that dislocations 
will appear at some length scale $R_{d}$ was 
made in the context of equilibrium systems. 
In the moving vortex case it may still be possible to obtain 
a defect free lattice
in two dimensions as suggested by Koshelev and Vinokur \cite{Koshelev}.  

\begin{figure}
\includegraphics[width=3.5in]{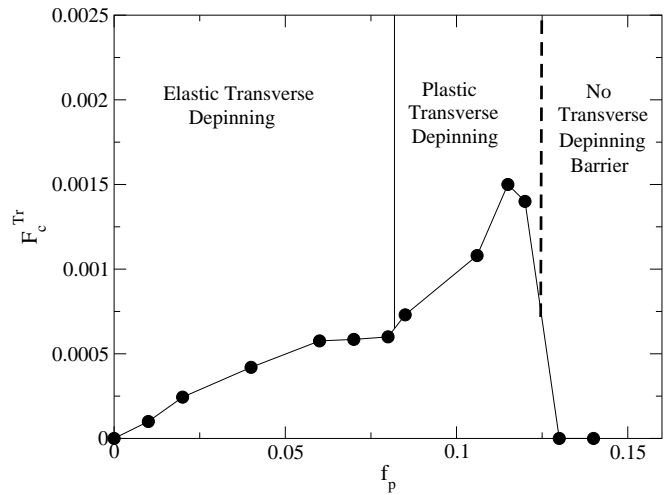}
\caption{
$F_c^{Tr}$ versus $f_p$ for $F^L_d=0.012f_0$ showing 
the three different response
regimes.  For weak pinning, $f_p<0.085f_0$, the transverse depinning
is elastic.  For intermediate pinning, $0.085 \le f_p/f_0 < 0.13$,
the transverse depinning is plastic.  There is no transverse depinning
barrier for strong pinning $f_p\ge 0.13f_0$.
}
\label{fig:new}
\end{figure}

In Fig.~5 we plot the transverse depinning barrier 
$F_c^{Tr}$ vs $f_{p}$ and indicate the three different pinning regimes
which can be distinguished by the nature of the transverse response.
In the weak pinning regime for $f_{p}/f_0 < 0.085$, 
the transverse depinning is elastic and a devil's staircase forms
when $F_d^{Tr}$ is swept.
Within this regime, $F^{Tr}_{c}$ increases monotonically with $f_p$. 
For the intermediate pinning regime $0.085 < f_{p}/f_0 < 0.13$, 
the transverse depinning is plastic and
$F^{Tr}_{c}$ is much higher than it was in the 
elastic depinning regime. 
For strong pinning $f_{p}/f_0 \ge 0.13$, 
the vortex lattice does not reorder when
the longitudinal drive is applied, and the transverse depinning
barrier disappears.

\begin{figure}
\includegraphics[width=3.5in]{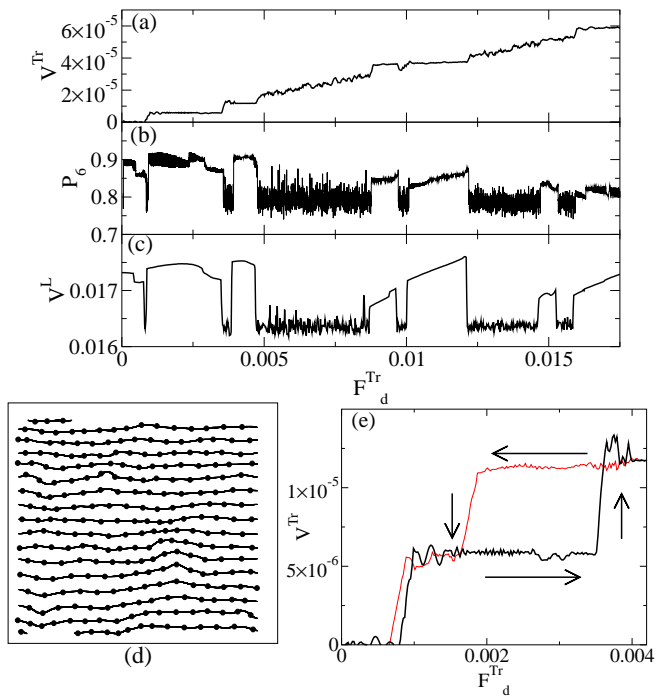}
\caption{
(a)$V^{Tr}$ vs $F^{Tr}_{d}$ for a sliding Wigner crystal with
fixed $F^L_d=0.03$.
(b) $P_{6}$ vs $F^{Tr}_{d}$. (c) $V^{L}$ vs $F^{Tr}_{d}$.
(d) The particle trajectories 
for the transversely pinned sliding state at $F^{Tr}_{d} = 3\times 10^{-4}$.
(e) The hysteresis in the $V^{Tr}$ vs $F^{Tr}_{d}$ curve.  Lower curve:
increasing $F^{Tr}_d$. Upper curve: 
decreasing $F^{Tr}_d$.
}
\label{fig:phase}
\end{figure}

\section{Transverse Depinning For Sliding Wigner Crystals on Smooth Potential Landscapes} 

We now consider the effect of a smoothly varying potential substrate 
of the type that arises in Wigner crystal systems for charge disorder.
In general, with this type of disorder we  
observe much larger relative critical transverse depinning thresholds
$F^{Tr}_c/F^L_c$
than with short range pinning. 
For long range disorder $F^{Tr}_{c}/F^{L}_{c}$  can be as high as $0.4$, while
for short range disorder it is less than $0.1$ to $0.01$. 
The effective shaking temperature $T^{sh}$, 
which arises from the rapidly fluctuating force caused by the pinning, 
for particles 
moving over short range random disorder at velocity $V$ is
$T^{sh} \propto 1/V$ \cite{Koshelev}. 
At high drives, the effective temperature drops with increasing $V$,
and the system can reorder.  At lower drives, $V$ is smaller and the
system melts into the plastic flow phase. 
In the smectic phase, the transverse depinning barrier
increases as the longitudinal drive is decreased \cite{Fangohr} 
so that the maximum transverse barrier occurs
just before the longitudinal drive is so low that the
shaking temperature melts the lattice and $F^{Tr}_{d}$ is lost. For 
long range, smoothly disordered substrates, 
there is no rapidly fluctuating force so the concept of an 
effective temperature does not apply. 
This means that a partially ordered moving state can 
persist to drives much closer to the 
longitudinal depinning threshold than in the case of short range pinning. 

In Fig.~\ref{fig:phase}(d), the particle 
trajectories for the case of long range charge disorder 
show that the channels are much more meandering than in 
Fig.~\ref{fig:delaunay}(b).      
In Fig.~\ref{fig:phase}(a,b,c) we plot
$V^{Tr}$, $P_{6}$, and the longitudinal velocity 
$V^{L}$ for the sliding Wigner crystal system  
with fixed $F^{L} = 0.03$. 
In this case the system initially forms a 
moving smectic system with $P_6=0.9$. A  step structure 
appears in $V^{Tr}$.  On the step,  
the particles are flowing in channels 
and $P_{6}$ shows only small fluctuations. 
In regions where $V^{Tr}$ is linearly increasing, 
$P_{6}$ rapidly fluctuates around a lower value,
indicating that the system is more disordered. 
The order-disorder transitions also affect 
$V^{L}$, which decreases and shows strong fluctuations
in the disordered regimes. In the ordered regimes, there are few fluctuations
in $V^{L}$ and there is a trend for $V^{L}$ to increase with $F^{Tr}_d$. 
This effect is similar to loading-unloading
cycles of the type found in friction systems. 
In Fig~\ref{fig:phase}(e) we illustrate the presence of hysteresis 
in $V^{Tr}$ vs $F^{Tr}_{d}$, which gives evidence that 
the order-disorder transitions have a first order characteristic.

\section{Summary} 

In summary, using numerical simulations we have confirmed the 
prediction of Giamarchi and Le Doussal
that a devil's staircase transverse response occurs
for an elastic vortex lattice moving over random pinning 
when the net force vector aligns with 
the symmetry directions of the moving lattice.
The devil's staircase response appears 
in the weak pinning regime when both the equilibrium state 
and the moving state are free of dislocations
and the transverse depinning occurs elastically 
without the generation of defects. 
For intermediate pinning strengths, where the equilibrium state is
dislocated and the vortices form a moving smectic state under a longitudinal
drive, the transverse
depinning occurs in a series of hysteretic 
order-disorder transitions where the system alternately
becomes disordered and then 
reorganizes into channel structures 
that exhibit a transverse depinning threshold. In this case a
series of pronounced steps in the transverse response appear. 
In the case where the quenched disorder
is long-range and smoothly varying, 
the order-disorder transitions dominate the response to
an increasing applied 
transverse force.  These results are applicable to vortices and 
Wigner crystals
moving over random disorder. 

This work was supported by the U.S. Department of Energy under Contract
No. W-7405-ENG-36.

\end{document}